\documentclass{jpsj-suppl}

\title{Gauge fields, massless modes and topology of gauge
fields in multi-band superconductors
}

\author{Takashi \textsc{Yanagisawa}, Yasumoto \textsc{Tanaka} and 
Izumi \textsc{Hase}
}

\inst{Electronics and Photonics Research Institute,
National Institute of Advanced Industrial Science and Technology (AIST),
Tsukuba Central 2, 1-1-1 Umezono, Tsukuba 305-8568, Japan
}

\email{t-yanagisawa@aist.go.jp}

\recdate{July 14, 2013}

\abst{
Multi-phase physics is a new physics of multi-gap superconductors.
Multi-band superconductors exhibit many interesting and novel properties.
We investigate the dynamics of the phase-difference mode and show that
this mode yields a new excitation mode.
The phase-difference mode is represented as an abelian vector field.
There are massless modes when the number of gaps is greater than three
and the Josephson term is frustrated.
The fluctuation of phase-difference modes with non-trivial topology 
leads to the existence of a
fractional-quantum flux vortex in a magnetic field.
A superconductor with a fractional-quantum flux vortex is regarded as a
topological superconductor with the integer Chern number.
}
 
\kword{multi-band superconductor, phase-difference mode, fractional vortex, 
massless mode}

\begin{document}
\maketitle

\section{Introduction}
There are interesting and profound analogies between particle physics 
and superconductivity.
This was first pointed out by Y. Nambu, and he invented a concept
of spontaneous symmetry breaking in particles  
physics\cite{nam61a,nam60b}.
The global $U(1)$ phase invariance is spontaneously broken in
superconductors.
It is well known that the gapless Goldstone mode exists when the
continuous symmetry is spontaneously broken.  
Superconductivity is most familiar phenomenon that occurs as a
result of spontaneous symmetry breaking.
The Ginzburg-Landau free energy describes a spontaneous breaking 
of $U(1)$ symmetry.
The order parameter is written as $\psi= |\psi|e^{i\theta}$
for any real angle $\theta$ in the range $0\leq\theta\leq 2\pi$.
Any choice of $\theta$ would have exactly the same energy that implies
the existence of a massless Nambu-Goldstone boson.
This changes qualitatively when the Coulomb interaction between the electrons
is included.  The Coulomb repulsive interaction turns the massless
mode into a gapped plasma mode\cite{and58}.
Therefore the mode that originates from the phase variable $\theta$ does
not play an important role in single-band superconductors.
This would change qualitatively again in multi-gap superconductors because
the multi-phase mode variables will
produce new excition states.  

Multi-phase physics is a new physics of multi-gap superconductors.
The study of multi-gap superconductors stemmed from works by Kondo\cite{kon63}
and Suhl et al.\cite{suh59}.
An additional phase invariance will bring about novel phenomena.
The phase-difference mode would yield new 
phenomena\cite{leg66,gei67,agt99,sha02,gur03} and new excitation
modes in multi-gapped superconductors.
The negative isotope effect in Fe pnictides is an example of the multi-band
effect\cite{shi09,yan09}. 
The existence of fractionally quantized flux vortices is very
significant and attractive.  The fluctuation of phase-difference mode leads to
half-quantum flux vortices in two-gap superconductors\cite{izu90,tan02,bab02}.
A generalization to a three-gap superconductor results in very
attractive features, that is,  chiral states with time-reversal
symmetry breaking and the existence of fractionally quantized
vortices\cite{sta10,tan10a,tan10b,yan12,lin12,nit12,pla12}.


\section{Gauge fields and the free energy}


Superconductivity is phenomenologically described by the Ginzburg-Landau
free energy.
We first consider the Ginzburg-Landau free energy density of a two-band
superconductor without the Josephson term in a magnetic field:
\begin{eqnarray}
f&=& (\alpha_1|\psi_1|^2+\alpha_2|\psi_2|^2)+\frac{1}{2}(\beta_1|\psi_1|^4
+\beta_2|\psi_2|^4)\nonumber\\
&+& \frac{\hbar^2}{2m_1}\Big|\left(\nabla-i\frac{e^*}{\hbar c}{\bf A}\right)
\psi_1\Big|^2
+ \frac{\hbar^2}{2m_2}\Big|\left(\nabla-i\frac{e^*}{\hbar c}{\bf A}\right)
\psi_2\Big|^2
+ \frac{1}{8\pi}(\nabla\times{\bf A})^2,
\end{eqnarray}
where $\psi_j$ $(j=1,2)$ are the order parameters and $e^*=2e$.
This functional is not invariant under the transformation:
\begin{equation}
\psi_j\rightarrow \exp\left(i\frac{e^*}{\hbar c}\bar{\theta_j}\right)\psi_j,~~~
{\bf A}\rightarrow {\bf A}+\nabla\chi.
\end{equation}
The functional is not invariant for any choice of $\chi$.
Let us assume that the phase of $\psi_j$ is $\theta_j$:
$\psi_j=e^{i\theta_j}\rho_j$, and define $\Phi=\theta_1+\theta_2$ and
$\varphi=\theta_1-\theta_2$, where $\rho_j=|\psi_j|$. 
The free energy is written as
\begin{eqnarray}
f&=& (\alpha_1|\rho_1|^2+\alpha_2|\rho_2|^2)+\frac{1}{2}(\beta_1|\rho_1|^4
+\beta_2|\rho_2|^4)\nonumber\\
&+& \frac{\hbar^2}{2m_1}\Big|\left(\nabla-i\frac{e^*}{\hbar c}{\bf A}
-i\frac{e^*}{\hbar c}{\bf B}\right)\rho_1\Big|^2
+ \frac{\hbar^2}{2m_2}\Big|\left(\nabla-i\frac{e^*}{\hbar c}{\bf A}
+i\frac{e^*}{\hbar c}{\bf B}\right)\rho_2\Big|^2
+\frac{1}{8\pi}(\nabla\times{\bf A})^2,\nonumber\\
\end{eqnarray}
where
\begin{equation}
{\bf B}= -\frac{\hbar c}{2e^*}\nabla\varphi,
\end{equation}
and we write ${\bf A}-\hbar c/(2e^*)\nabla\Phi$ as ${\bf A}$.


It is straightforward to generalize the free energy to an $N$-band
superconductor.
In this case, we have $N-1$ phase-difference modes.
$N-1$ equals the rank of $SU(N)$.  The rank is the number of elements
of Cartan subalgebra, namely commutative generators.
Let $t_1,\cdots,t_{N-1}$ be elements of the
Cartan subalgebra of $SU(N)$.  Then, the covariant derivative is
\begin{equation}
D_{\mu}= \partial_{\mu}-i\frac{e^*}{\hbar c}A_{\mu}
-i\frac{e^*}{\hbar c}\sum_{j=1}^{N-1}B_{\mu}^jt_j,
\end{equation}
and the free energy density (without the Josephson terms) is given by
\begin{equation}
f= \sum_j\alpha_j|\rho_j|^2+\frac{1}{2}\sum_j\beta_j|\rho_j|^4
+\frac{\hbar^2}{2m}|D_{\mu}\psi|^2+\frac{1}{8\pi}
(\nabla\times{\bf A})^2.
\end{equation}
Here, we adopted that masses are the same and
$\psi=(\rho_1,\cdots,\rho_N)^t$ is a scalar field of order
parameters.
The phase-difference modes $B_{\mu}^j$ are represented by the diagonal part
of $SU(N)$ nonabelian gauge fields and correspond to the abelian
projection of $SU(N)$ gauge theory by 'tHooft\cite{tho81}.

\section{Josephson term and massless modes}

There are $N-1$ gauge fields $B_{\mu}$ in the $N$-gap
superconductors.  We add the Josephson term to the free energy
functional, representing the pair transfer interactions between
different conduction bands\cite{kon63}.
The Josephson term is given as
$V = -\sum_{i\neq j}\gamma_{ij}|\psi_i||\psi_j|\cos(\theta_i-\theta_j)$,
where $\gamma_{ij}=\gamma_{ji}$ are chosen real.
This term obviously loses the gauge invariance of the free energy or
the Lagrangian because $\theta_i-\theta_j$
is not gauge invariant.
This indicates that the phase-difference modes acquire masses.
Hence, in the presence of the Josephson term, the phase-difference modes
are massive and there are excitation gaps.

This would change qualitatively when $N$ is greater than 3\cite{tan11,yan13}.
We show that massless modes exist for an $N$-equivalent frustrated band
superconductor. 
Let us consider the Josephson potential given by
\begin{equation}
V= \Gamma [ \cos(\theta_1-\theta_2)+\cos(\theta_1-\theta_3)
+\cos(\theta_1-\theta_4) 
+ \cos(\theta_2-\theta_3)+\cos(\theta_2-\theta_4)
+\cos(\theta_3-\theta_4) ],
\end{equation}
for $N=4$.  We assume that $\Gamma$ is positive: $\Gamma >0$ which indicates
that there is a frustration effect between Josephson couplings.
The ground states of this potential are degenerate.  For example,
the states with $(\theta_1,\theta_2,\theta_3,\theta_4)= (0,\pi/2,\pi,3\pi/2)$
and $(0,\pi,0,\pi)$ have the same energy.
The Fig.1 shows $V$ as a function of $\theta_1-\theta_3$ and
$\theta_2-\theta_4$ in the case of $\theta_1-\theta_3=\theta_2-\theta_4$.
By expanding $V$ around a minimum $(0,\pi/2,\pi,3\pi/2)$,
 we find that there is one massless
mode and two massive modes.
In fact, for $\theta_1-\theta_2=-\pi+\tilde{\eta}_1$,
$\theta_2-\theta_4=-\pi+\tilde{\eta}_2$ and
$\theta_2-\theta_3=-\pi/2+\tilde{\eta}_3$,
the potential $V$ is written as
$V= \Gamma [ -2+(1/2)\tilde{\eta}_1^2+(1/2)\tilde{\eta}_2^2+\cdots ]$,
where the dots indicate higher order terms.
Missing of $\tilde{\eta}_3^2$ means that there is a massless mode
and there remains a global $U(1)$ rotational symmetry, indicating that
 the ground states are continuously degenerate.
The gauge field corresponding to $\theta_2-\theta_3$ represents a massless
mode near $(\theta_1,\theta_2,\theta_3,\theta_4)=(0,\pi/2,\pi,3\pi/2)$.
One gauge symmetry is not broken and two gauge symmetries are broken for $N=4$.
The massive modes are represented by
linear combinations of $\theta_1-\theta_3$ and $\theta_2-\theta_4$.
When we expand the potential $V$ near the minimum
$(\theta_1,\theta_2,\theta_3,\theta_4)=(0,\pi,0,\pi)$, we obtain
$V= \Gamma [ -2+\frac{1}{2}\eta_1^2+\cdots ]$.
This indicates that there are two massless modes and one massive mode
(see Fig.2).

\begin{figure}[htbp]
\begin{tabular}{cc}
\begin{minipage}{0.5\hsize}
\begin{center}
\includegraphics[width=4.5cm,angle=90]{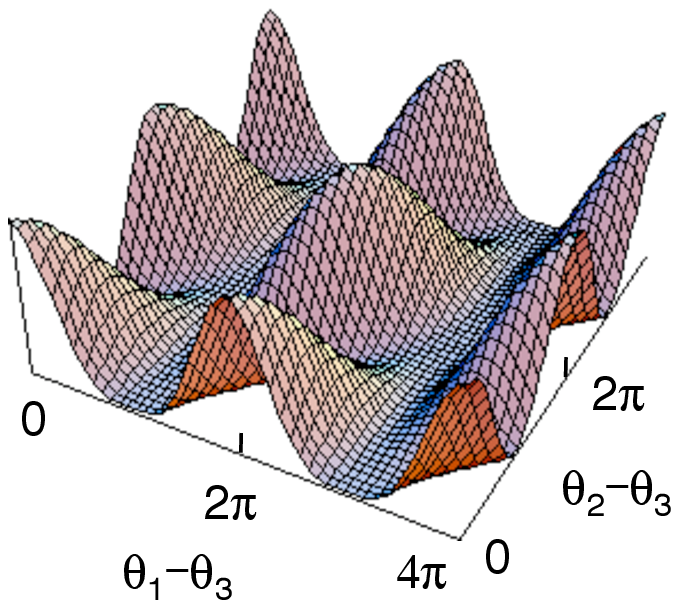}
\caption{
Josephson potential for the 4-band band as a function of
$\theta_1-\theta_3$ and $\theta_2-\theta_3$.
We set $\theta_1-\theta_3=\theta_2-\theta_4$ in the potential.
The flat minimum indicates an existence of zero mode.
}
\end{center}
\end{minipage}
\begin{minipage}{0.5\hsize}
\begin{center}
\includegraphics[width=3.2cm,angle=90]{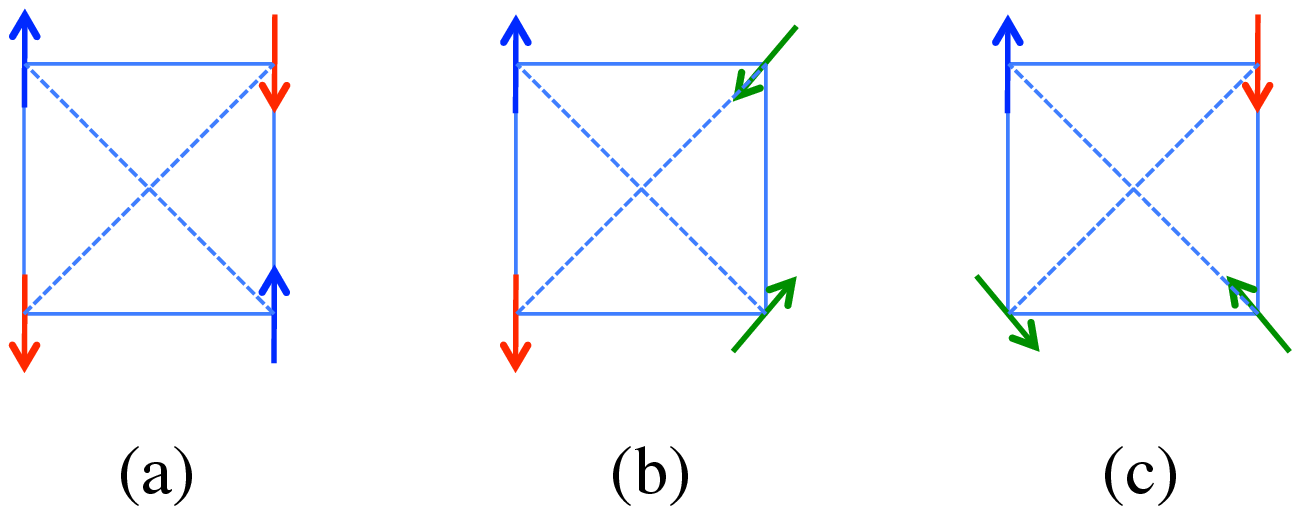}
\caption{
Configurations which have the same energy where angles $\theta_j$ are
shown by arrows.
In (b) and (c) two spins can be rotated with the phase difference
fixed to be $\pi$ keeping the energy constant.
}
\end{center}
\end{minipage}
\end{tabular}
\end{figure}

We can generalize this argument for general $N$.  We show that
for $N\ge 4$, there exist always the massless modes for the potential
\begin{equation}
V= \Gamma [ \cos(\theta_1-\theta_2)+\cos(\theta_1-\theta_3)+\cdots
+\cos(\theta_1-\theta_N)+\cdots+\cos(\theta_{N-1}-\theta_N) ].
\label{v-n}
\end{equation}
For $\Gamma>0$, there are two massive modes and $N-3$ massless modes,
near the minimum
$(\theta_1,\theta_2,\theta_3,\theta_4,\cdots)=(0,2\pi/N,4\pi/N,6\pi/N,\cdots)$.
Near the minimum $(\theta_1,\theta_2,\theta_3,\cdots)=(0.\pi,0,\cdots)$, 
we have $N-2$ massless modes and one
massive mode.

\section{Non-trivial configuration of gauge fields}

The phase-difference gauge field ${\bf B}$ in the two-gap case is defined as
${\bf B}= -(\hbar c/(2e^*))\nabla\varphi$.
The half-quantum vortex can be interpreted as a monopole.
Let us assume that there is a cut, namely, kink on the real axis
for $x>0$.  The phase $\theta_1$ is represented by
$\theta_1 = -\frac{1}{2}{\rm Im}\log\zeta+\pi$,
where
$\zeta= x+iy$.
The singularity of $\theta_j$ can be transferred to a singularity of the
gauge field by a gauge transformation.
We consider the case $\theta_2=-\theta_1$: $\phi=2\theta_1$.  Then we have
\begin{equation}
{\bf B}= -\frac{\hbar c}{2e^*}\nabla\phi= -\frac{\hbar c}{e^*}\frac{1}{2}
\left( \frac{y}{x^2+y^2},-\frac{x}{x^2+y^2},0 \right).
\end{equation}
Thus, when the gauge field ${\bf B}$ has a monopole-type singularity,
the vortex with half-quantum flux exists in two-gap superconductors.
The one-form corresponding to ${\bf B}$ defines the Chern class and the
integral of it over the sphere $S^2$ gives the Chern number $C_1$.
In general, the gauge field ${\bf B}$ has the integer Chern number:
$C_1=n$.  For $n$ odd, we have a half-quantum flux vortex.



\begin{thebibliography}{}

\bibitem{nam61a}Y. Nambu and G. Jona-Lasinio: Phys. Rev. 122 (1961) 345;
124 (1961) 246..
\bibitem{nam60b}Y. Nambu: Phys. Rev. 117 (1960) 648.
\bibitem{and58}P. W. Anderson: Phys. Rev. 112 (1958) 1900.
\bibitem{kon63}J. Kondo: Prog. Theor. Phys. 29 (1963) 1.
\bibitem{suh59}H. Suhl, B. T. Mattis and L. W. Walker: Phys. Rev. Lett.
3 (1959) 552.
\bibitem{leg66}A. J. Leggett: Prog. Theor. Phys. 36 (1966) 901.
\bibitem{gei67}B. T. Geilikman, R. O. Zaitsev and V. Z. Kresin: Soviet Phys.-
Solid State 9 (1967) 642.
\bibitem{agt99}D. F. Agterberg, V. Barzykin and L. P. Gorkov: Phys. Rev.
B60 (1999) 14868.
\bibitem{sha02}S. C. Sharapov, V. P. Grusynin and H, Beck: Eur. Phys. J.
B30 (2002) 45.
\bibitem{gur03}A. Gurevich: Phys. Rev. B67 (2003) 184515.
\bibitem{shi09}P.M. Shirage et al.:
Phys. Rev. Lett. 103 (2009) 257003.
\bibitem{yan09}T. Yanagisawaa et al.:
J. Phys. Soc. Jpn. 78 (2009) 094718; ibid. 79 (2010) 126002.
\bibitem{izu90}Yu. A. Izyumov and V. M. Laptev: Phase Transitions 20 (1990) 95.
\bibitem{tan02}Y. Tanaka: Phys. Rev. Lett. 88 (2002) 017002.
\bibitem{bab02}E. Babaev: Phys. Rev. Lett. 89 (2002) 067001.
\bibitem{sta10}V. Stanev and Z. Tesanovic: Phys. Rev. B81 (2010) 134522.
\bibitem{tan10a}Y. Tanaka and T. Yanagisawa: J. Phys. Soc. Jpn. 79 (2010) 114706.
\bibitem{tan10b}Y. Tanaka and T. Yanagisawa: Solid State Commun. 150 (2010) 1980.
\bibitem{yan12}T. Yanagisawa, Y. Tanaka, I. Hase and K. Yamaji: J. Phys. Soc. Jpn.
81 (2012) 024712.

\bibitem{lin12}S. Z. Lin and X. Hu: New J. Phys. 14 (2012) 063021.
\bibitem{nit12}M. Nitta, M. Eto, T. Fujimori and K. Ohashi: J. Phys. Soc. Jpn.
81 (2012) 084711.
\bibitem{pla12}C. Platt, R. Thormale, C. Honerkamp and S. C. Zhang:
Phys. Rev. B85 (2012) 180502.
\bibitem{tho81}G. 't Hooft: Nucl. Phys. 190 (1981) 455.
\bibitem{tan11}Y. Tanaka et al.: Physica C471 (2011) 747.
\bibitem{yan13}T. Yanagisawa and I. Hase: A proof will be shown in future;
submitted to J. Phys. Soc. Jpn.

\end{thebibliography}
\end{document}